\documentstyle[12pt]{article}

\begin{document}

\title{The Central Term in 3D Simple Superalgebra}
\author{L. P. Colatto\thanks{%
e-mail: colatto@fis.unb.br} \\
NRTP - Instituto de F\'\i sica - UnB - DF - CEP 70919-900 -\ \ Brazil.}
\date{}
\maketitle

\begin{abstract}
A matter self-interacting model with $N=1$-supersymmetry in 3D is discussed
in connection with the appearance of a central charge in the algebra of the
supersymmetry generators. The result is extended to include gauge fields
with a Chern-Simons term. The main result is that, for a simple
supersymmetry, only the matter sector contributes to the central charge in
contrast to what occurs in the $N=2$ case.
\end{abstract}

\smallskip 
\thispagestyle{empty}

\newpage \smallskip

Ordinary and supersymmetric Abelian gauge models in three-dimensional
space-times have been fairly-well investigated in varoius contexts over the
past years \cite{alguem}. Besides their relevance in connection with the
possibility of getting non-perturbative results more easily, the ultraviolet
finiteness of Yang-Mills (and gravity) Chern-Simons models is a remarkable
feature of field theories defined in $D=(1+2)$ \cite{piguet}. Also, $3D$
gauge theories seem to be the right way to tackle exciting topics of
Condensed Matter Physics such as High-$T_c$ Superconductivity and Fractional
Quantum Hall Effect \cite{alguemdois}.

Our purpose in this paper is to assess an Abelian three-dimensional gauge
model with $N=1$ supersymmetry, from the point-of-view of the algebra of
supersymmetry generators. We actually wish to present here a few remarks on
the connection between topologically non-trivial solutions, the Chern-Simons
term, and the presence of a central charge operator in the supersymmetry
algebra\cite{witten}.

The super-Poincar\'{e} algebra in $(1+2)$ dimensions is generated by a real
two-component spinorial charge, $Q_{a}$, whose operatorial relations are
listed below: 
\begin{equation}
\{Q_{a}\,,\,Q_{b}\}\,=\,2P_{ab}\;\;\;\mbox{e}\;\;\;[Q_{a}\,,\,P_{ab}]\,=\,0%
\,,  \label{algoper}
\end{equation}
where $P_{ab}$ is the translation generator. We shall represent vectors in a
twofold way: for Lorentz indices we will use greek letters, and for
bi-spinorial indices we will use latin letters, bearing in mind the mapping $%
V_{ab}=V_{\mu }(\gamma ^{\mu })_{ab}$ \cite{grisaru}. The super-Poincar\'{e}
algebra (\ref{algoper}) for an extended supersymmetry with $N$ flavours is
generalized to \cite{haag} 
\begin{eqnarray}
\{Q_{a}^{i}\,,\,Q_{b}^{j}\} &=&2\delta ^{i\,j}P_{ab}\,+\,A^{i\,j}\epsilon
_{ab}\,,  \nonumber \\
{\lbrack }Q_{a}^{i}\,,\,P_{ab}{]} &=&0\,,  \nonumber \\
{\lbrack }Q_{a}^{i}\,,\,A^{j\,k}{]} &=&0\,,  \label{algeger}
\end{eqnarray}
where $i,j,k\,=\,1,...,N$, $\epsilon _{ab}$ is the Levi-Civita tensor in
spinor space and $A^{i\,j}=-A^{j\,i}$ is the central charge that transforms
under the symmetry group which defines an automorphism of the algebra of
extended supersymmetry.

To find out if a quantum field theory exhibits supersymmetry, one might
study the set of Ward identities among the Green's functions, and then
establish whether or not they are respected. In our work, we adopt another
procedure, namely, we explicitly compute the equal-time current algebra
associated to supersymmetry; such a method is able to automatically signal
the eventual presence of central charge operators as originated from
topologically non-trivial field configurations \cite{witten}.

Adopting the metric tensor as $\eta^{\mu \, \nu} \, = \, (+ ; - , -)$, we
shall choose the following representation for the $\gamma$-matrices:

\begin{equation}
{(\gamma ^{0})^{a}}_{b}\,=\,\left( 
\begin{array}{cc}
0 & -i \\ 
i & 0
\end{array}
\right) \;,\;{(\gamma ^{1})^{a}}_{b}\,=\,\left( 
\begin{array}{cc}
0 & -i \\ 
-i & 0
\end{array}
\right) \;,\;{(\gamma ^{2})^{a}}_{b}\,=\,\left( 
\begin{array}{cc}
-i & 0 \\ 
0 & i
\end{array}
\right) \,,  \label{gama}
\end{equation}
\begin{equation}
C_{ab}\,=\,\left( 
\begin{array}{cc}
0 & -\,i \\ 
i & 0
\end{array}
\right) \,,
\end{equation}
where we have the ``metric'' tensor $C_{ab}\,=\,-C_{ba}\,=\,C^{ab}$ e $%
C_{ab}C^{cd}\,=\,{\delta ^{c}}_{[a}\,{\delta ^{d}}_{b]}$. We list below a
number of algebraic relations among the $\gamma $-matrices that have been
useful in our component-field calculations: 
\begin{eqnarray}
&(\gamma ^{\mu })^{ab}(\gamma ^{\nu })_{ab}\,=\,\eta ^{\mu \,\nu }\,,& 
\nonumber \\
&(\gamma ^{\mu })^{ab}(\gamma ^{\nu })_{bd}\,=\,\eta ^{\mu \,\nu }{\delta
^{a}}_{d}\,-\,i\,{\epsilon ^{\mu \,\nu }}_{\rho }{(\gamma ^{\rho })^{a}}%
_{d}\,,&  \nonumber \\
&{(\gamma ^{\mu })_{a}}^{b}{(\gamma ^{\nu })^{a}}_{b}\,=\,-\,\eta ^{\mu
\,\nu }\,,&  \nonumber \\
&(\gamma ^{\mu })_{ab}{(\gamma ^{\nu })^{b}}_{d}\,=\,-\,\eta ^{\mu \,\nu
}C_{da}\,-\,i\,\epsilon ^{\mu \,\nu \,\rho }(\gamma _{\rho })_{ad}\,,& 
\nonumber \\
&{(\gamma ^{\mu })_{a}}^{b}{(\gamma ^{\nu })_{b}}^{d}\,{(\gamma ^{\rho })_{d}%
}^{a}\,=\,-\,i\,\epsilon ^{\mu \,\nu \,\rho }\,,&  \nonumber \\
&(\gamma ^{\mu })_{ab}{(\gamma ^{\nu })^{b}}_{c}\,{(\gamma ^{\rho })^{c}}%
_{d}\,=\,i\,\epsilon ^{\mu \,\nu \,\rho }C_{da}\,+\,\eta ^{\rho \,\nu
}(\gamma ^{\mu })_{da}\,-\,\eta ^{\rho \,\mu }(\gamma ^{\nu })_{da}\,-\,\eta
^{\mu \,\nu }(\gamma ^{\rho })_{da}\,.&  \label{clifgama}
\end{eqnarray}

This paper is outlined as follows: in Section 1, a self-interacting scalar
model is presented and the su.sy. algebra is written down with the explicit
form for the central charge operator; the introduction of the gauge sector
is discussed in Section 2. Finally, in Section 3, one discusses the
supersymmetric version of a Chern-Simons term, and the connection it bears
with the central charge is investigated. Our General Conclusions follow in
Section 4.

%

\section{Self--Interacting Scalar Model and Vortex \newline Configurations}

%

The component-field expansion for a scalar superfield reads 
\begin{equation}
\Phi(x,\theta) \; = \; A(x) \, + \, \theta^a \, \psi_a(x) \, - \, \theta^2
\, F(x) \; ;
\end{equation}
where $\theta$ is a (real) Grassmann-valued Majorana spinor and $A(x)$ is a
physical scalar, $\psi_a(x)$ is a physical fermion and $F(x)$ is an
auxiliary field. The supersymmetry covariant derivative is represented as 
\begin{eqnarray}
D_a \; = \; \partial_a \, + \, i \, \theta^b \, \partial_{ab} \; , \\
{\{} D_a \, , \, D_b{\}} \, = \, 2 P_{ab} \, .
\end{eqnarray}

The most general $N$=$1$-supersymmetric action with renormalizable matter
self-inter\-actions is given by 
\begin{equation}
S_{scalar}\;=\;\int \,d^{3}x\,d^{2}\theta \left\{ \,-\frac{1}{2}%
\,(D_{a}\,\Phi )^{2}\,+\frac{1}{2}\,m\,\Phi ^{2}\,+\,\frac{\lambda }{8}%
\,\Phi ^{4}\,\right\} \;,  \label{acaosuper}
\end{equation}
and the supersymmetry transformations on the components $A$, $\psi _{a}$ and 
$F$ read as below: 
\begin{eqnarray}
\delta \,A\, &=&\,-\epsilon ^{a}\,\psi _{a}\,,  \nonumber \\
\delta \,\psi _{a}\, &=&\,-\epsilon ^{b}\,(C_{ab}\,F\,+\,i\partial _{ab}A)\,,
\nonumber \\
\delta \,F\, &=&\,-\epsilon ^{b}\,i\,{\partial _{b}}^{a}\,\psi _{a}\;.
\label{vardecampos}
\end{eqnarray}

The action for the physical fields, 
\begin{eqnarray}
S_{scalar}\, &=&\,\int \,d^{3}x\,\left\{ 
\begin{array}{l}
\frac{1}{2}\,\left[ \,\frac{1}{2}(\partial _{ab}\,A)(\partial
^{ab}\,A)\,+\,\psi ^{a}\,i\,{\partial _{a}}^{b}\,\psi _{b}\right]
\,+\,m\,\psi ^{2}\,+\,\frac{3}{2}\,\lambda \,\psi ^{2}\,A^{2}
\end{array}
\right. \,+\,  \nonumber \\
&&-\left. 
\begin{array}{r}
\frac{1}{2}\,m^{2}\,A^{2}\,-\,\frac{1}{2}\,m\,\lambda \,A^{4}\,-\,\frac{1}{8}%
\,\lambda ^{2}\,A^{6}\,
\end{array}
\right\} \,,  \label{acaoescal}
\end{eqnarray}
can be shown to be invariant under the non-linear ``on-shell''
transformations, 
\begin{eqnarray}
\delta \,A\, &=&\,-\epsilon ^{a}\,\psi _{a}\,,  \nonumber \\
\delta \,\psi _{a}\, &=&\,-\epsilon ^{b}\,\left[ C_{ab}\,\left( -\frac{1}{2}%
\,m\,A\,-\,\frac{1}{2}\lambda \,A^{3}\right) \,+\,i\partial _{ab}A\right] \,.
\label{transescal}
\end{eqnarray}

Now , taking into account that supersymmetry is a symmetry of the action
(the Lagrangian density transforms as a total derivative), it can be shown
that the Noether current associated to $N$=$1$-supersymmetry turns out to
be: 
\begin{eqnarray}
{J^{\mu}}_c & = & \, - \, i \, \psi^a \, (\gamma^{\mu})_{ac} \, \left( m \,
A + \, \frac{\lambda}{2} \, A^3 \right) \, - \, \frac{i}{2} \,
\varepsilon^{\mu \, \nu \, \rho} \, \psi^b \, (\gamma_{\rho})_{bc} \,
\partial_{\nu} \, A \, +  \nonumber \\
& & + \, \frac{1}{2} \, \psi_c \, \partial^{\mu} \, A \, + \, \frac{1}{2} \,
A \, \partial^{\mu} \, \psi_c \, - \, \frac{i}{2} \, \varepsilon^{\mu \, \nu
\, \rho} \, A \, \partial_{\nu} \, \psi^a \, (\gamma_{\rho})_{ac} \, .
\label{supercorr}
\end{eqnarray}
The supercharge is defined as 
\begin{eqnarray}
Q_c & = & \int \, d^2\vec{x} {J^0}_c  \nonumber \\
& = & \int \, d^2\vec{x} \left\{ -i \, \psi^a \, (\gamma^0)_{ac} \, \left( m
\, A + \, \frac{\lambda}{2} \, A^3 \right) \, + \, \frac{1}{2} \, \psi_c \,
\partial^0 \, A \, + \, \frac{1}{2} \, A \, \partial^0 \, \psi_c \, + \right.
\nonumber \\
& & \left. - \, \frac{i}{2} \, \varepsilon^{0 \, \nu \, \rho} \, \psi^a \,
(\gamma_{\rho})_{ac} \, \partial_{\nu} \, A \, + \, \frac{i}{2} \,
\varepsilon^{0 \, \nu \, \rho} \, A \, \partial_{\nu} \, \psi^a \,
(\gamma_{\rho})_{ac} \right\} \, .  \label{cargadesusy}
\end{eqnarray}

With the help of the canonical commutation (and anticommutation) relations
for the physical fields, a tedious calculation yields the following
expression for the algebra of supersymmetry charges: 
\begin{eqnarray}
&&\{Q_{a}\,,\,Q_{b}\}\,=\,\int \,d^{2}\vec{x}\times %
\mbox{~~~~~~~~~~~~~~~~~~~~~~~~~~~~~~~~~~~~~~~~~~~~~~~~~~~~~~~~~~~~~~~~~~~~} 
\nonumber \\
&&-2\,i\left\{ \frac{1}{4}\left[ 
\begin{array}{l}
2\,i\ \psi ^{a}\,{(\gamma ^{0})_{a}}^{b}\,\partial ^{0}\,\psi
_{b}\,+\,(\partial ^{0}\,A)(\partial ^{0}\,A)\,+\,2\,i\,\psi ^{a}\,{(\gamma
^{i})_{a}}^{b}\,\partial ^{i}\,\psi _{b}\,+(\partial ^{i}\,A)(\partial
^{i}\,A)\,+
\end{array}
\right. \right.  \nonumber \\
&&\,\,\,-\,\left. 
\begin{array}{l}
\left. 
\begin{array}{l}
\psi ^{2}\left( m\,+\,\frac{3}{2}\,\lambda \,A^{2}\right) \,+\,\left( \frac{1%
}{2}\,m^{2}\,A^{2}\,+\,\frac{1}{2}\,m\,\lambda \,A^{4}\,+\,\frac{1}{4}%
\,\lambda ^{2}\,A^{6}\right)
\end{array}
\right]
\end{array}
\right\} (\gamma ^{0})_{ab}\,+  \nonumber \\
&&-\,2\,i\left\{ \frac{1}{4} 
\begin{array}{l}
2\,i\,\psi ^{a}\,{(\gamma ^{(0})_{a}}^{b}\,\partial ^{i)}\,\psi
_{b}\,+\,(\partial ^{(0}\,A)\,(\partial ^{i)}\,A)
\end{array}
\right\} \,(\gamma _{i})_{ab}\,.  \label{algscescal}
\end{eqnarray}
If we compare this expression with the $0\,\mu $ component of the
``improved'' energy-momen\-tum tensor 
\begin{equation}
T_{\mu \,\nu }\,\equiv \,\frac{1}{e}\,\frac{\delta S}{\delta e_{a^{\prime
}}^{(\mu }}\,e_{\nu )\,a^{\prime }}\,=\,-\left. \frac{2}{\sqrt{-g}}\,\frac{%
\delta S}{\delta g^{\mu \,\nu }}\right| _{g^{\mu \nu }=\eta ^{\mu \nu }}\,,
\label{tmini}
\end{equation}
where $S$ in this expression indicates the action (\ref{acaoescal}), we may
rewrite (\ref{algscescal}) as 
\begin{equation}
\{Q_{a}\,,\,Q_{b}\}\,=\,2\,i\,P^{\mu }\,(\gamma _{\mu
})_{ab}\,+\,2\,i\,\epsilon ^{i\,j}\,\int \,d^{2}\vec{x}\,(\partial
_{i}\,A)\,(\gamma _{j})_{ab}\,\left( m\,A\,+\,\frac{\lambda }{2}A^{3}\right)
\,.  \label{mattercharge}
\end{equation}
In terms of the chiral components of the supersymmetry charge, the algebra
takes over the form: 
\begin{eqnarray}
\{Q^{+}\,,\,Q^{+}\} &=&2\,i\,(P^{0}\,+\,P^{1})\,-\,2\,\int d^{2}\vec{x}%
\,\left( mA\,+\,\frac{\lambda }{2}A^{3}\right) \partial _{2}A\,, \\
\{Q^{-}\,,\,Q^{-}\} &=&2\,i\,(P^{0}\,-\,P^{1})\,-\,2\,\int d^{2}\vec{x}%
\,\left( mA\,+\,\frac{\lambda }{2}A^{3}\right) \partial _{2}A\,,  \nonumber
\\
\{Q^{+}\,,\,Q^{-}\} &=&-2\,i\,P^{2}\,-\,2\,\int d^{2}x\,\left( mA\,+\,\frac{%
\lambda }{2}A^{3}\right) \partial _{1}A\,.  \label{mattercharge2}
\end{eqnarray}

Expressions (\ref{mattercharge}) and (\ref{mattercharge2}) sign the presence
of a central charge that is sensitive to a topologically non-trivial
behavior of the scalar sector at infinity: 
\begin{eqnarray}
T_{2} &=&\int dx_{1}\int dx_{2}\,\frac{\partial }{\partial _{x^{2}}}\left(
m\,A^{2}+\frac{\lambda }{4}\,A^{4}\right) \;,  \nonumber \\
T_{1} &=&\int dx_{2}\int dx_{1}\,\frac{\partial }{\partial _{x^{1}}}\left(
m\,A^{2}+\frac{\lambda }{4}\,A^{4}\right) \;,
\end{eqnarray}
where we observe the topological character of the central charge terms,
which has its origin in the mass and self-interacting terms of the scalar
field in Lagrangian. Bearing in mind this result, we shall now consider the
introduction of an Abelian gauge field with $N$=$1$--supersymmetry. We also
know that such a coupling is fundamental to stabilize the soliton-like
solutions in the form of magnetic vortices with finite energy.

\section{On the N=1 Super--QED$_3$}

The $N$=$1$--supersymmetric version of QED$_3$ is achieved upon the
complexification of the scalar superfield in eq. (\ref{acaosuper}) and the
gauge--covariantization of the spinor derivative: 
\begin{equation}
\nabla_a \, \equiv \, D_a \, \mp \, i \, \Gamma_a \, ,
\end{equation}
where $\Gamma_a$ is a gauge superconnection with super--helicity $h \, = \, 
\frac{1}{2}$, and the signs $-$ and $+$ indicate that the derivative is
acting in the superfields $\Phi$ and $\bar{\Phi}$ respectively ($\bar{\Phi}
\equiv \Phi^*$). $\Gamma_a$ admits the following $\theta$--expansion: 
\begin{equation}
\Gamma_a \, = \, \chi_a + \theta^b(C_{ab} B + i V_{ab}) + \theta^2(2
\lambda_a - i{\partial_a}^b \chi_b)\, ,
\end{equation}
where $\lambda_a$ is the gaugino field, $V_{ab}$ is the usual gauge field; $%
B $ and $\chi_a$ are compensating component--fields. In the so--called
Wess--Zumino gauge \cite{grisaru}, $\Gamma_a$ is reduced to: 
\begin{equation}
\Gamma_a = i \, \theta^b V_{ab} - 2 \, \theta^2 \, \lambda_a \, ,
\end{equation}
where the supersymmetry transformations read: 
\begin{eqnarray}
&\delta V_{ab} \, = \, i \, \epsilon_{(b} \, \lambda_{b)} \, ,&  \nonumber \\
&\delta \lambda_a \, = \, \frac{1}{2} \, \epsilon^c \, \partial_{c(a} \,
V_{c)b} \, .&
\end{eqnarray}

The covariantized vectorial derivative is written as 
\begin{equation}
\nabla _{ab}\,=\,D_{ab}\,\mp \,i\Gamma _{ab}\,,
\end{equation}
where $\Gamma _{ab}$ is the vector superconnection. As we know, in order to
have irreducible representations of symmetry, we need constraints in the
model. In the supersymmetric case, we have the so--called conventional
constraint, that acts in such away that the supersymmetric algebra of the
spinor derivatives, $\{\nabla _{a}\,,\,\nabla _{b}\}\,=\,2\,i\,\nabla
_{ab}\,+\,F_{ab}$, will have $F_{ab}\,=\,0$. Then, we easily compute that 
\begin{equation}
\Gamma _{ab}\,=\,-\frac{i}{2}\,D_{(a}\,\Gamma _{b)}\,,
\end{equation}
implying that in the {\it Wess--Zumino gauge} we have 
\begin{equation}
\Gamma _{ab}\,=\,V_{ab}\,+\,i\,\theta _{(a}\,\lambda _{b)}\,-\,\frac{i}{2}%
\,\theta ^{2}\,\partial _{c(a}\,{V_{b)}}^{c}\,.
\end{equation}
By the graded Bianchi identity, we redefine the gauge field as 
\begin{equation}
W_{a}\,=\,\frac{1}{2}\,D^{b}D_{a}\Gamma _{b}\,,
\end{equation}
with the constraint $D^{a}W_{a}\,=\,0$ ($D^{a}D_{b}D_{a}\,=\,0$), implying
as in the usual Lorentz gauge that exists only one independent component of
the field $W_{a}$. Using the projector method, we write 
\begin{equation}
W_{a}|\,=\,\lambda _{a}\;\;\;\;,\;\;\;\;D_{a}W_{b}|\,=\,\frac{1}{2}%
\,(\partial _{ca}\,{V_{b}}^{c}\,+\,\partial _{cb}\,{V_{a}}^{c})\,\equiv
\,f_{ab}\,,
\end{equation}
with $f_{ab}$ the usual gauge field strength. Another relations that will be
very important and that may straightforwardly be obtained are (conf. (\cite
{grisaru})): 
\begin{equation}
\nabla _{a}\,\nabla ^{2}\,=\,i\,{\nabla _{a}}^{b}\,\nabla _{b}\,\pm
\,i\,W_{a}\;\;\;\mbox{e}\;\;\;(\nabla ^{2})^{2}\,=\,\Box \,\mp
\,i\,W^{a}\,\nabla _{a}\,,
\end{equation}
where $\Box $ means the gauge--covariant d'Alembertian. Now, we are ready to
discuss the supersymmetry algebra in the framework of $N$=$1$ Super-QED$_{3}$%
.

\subsection{Scalar Superaction with a Background Gauge Field}

The U(1)--invariant superfield action without kinetic term for the gauge
sector is given as below: 
\begin{equation}
S_{scalar}\;=\;-\frac{1}{2}\int \,d^{3}x\,d^{2}\theta \left\{ (\nabla ^{a}\,%
\bar{\Phi})(\nabla _{a}\Phi )\right\} \,,  \label{acaoscalgauge}
\end{equation}
Redefining the component field, (we can always do this) by the projections 
\begin{eqnarray}
\Phi |=A\;\;\;\; &,&\;\;\;\;\bar{\Phi}|=\bar{A}\,,  \nonumber \\
\nabla _{a}\Phi |=\psi _{a}\;\;\;\; &,&\;\;\;\;\nabla _{a}\bar{\Phi}|=\bar{%
\psi}_{a}\,,  \nonumber \\
\nabla ^{2}\Phi |=F\;\;\;\; &,&\;\;\;\;\nabla \bar{\Phi}|=\bar{F}\,,
\end{eqnarray}
the gauge--field component action takes the form: 
\begin{equation}
S=\frac{1}{2}\,\int d^{3}x\left\{ \bar{\psi}^{a}\,i\,{D_{a}}^{b}\,\psi
_{b}\,+\,\psi ^{a}\,i\,{D_{a}}^{b}\,\bar{\psi}_{b}\,+\,A\,\Box \,\bar{A}\,+\,%
\bar{A}\,\Box A\right\} \,,  \label{acaoscalgau}
\end{equation}
with the ``on shell'' supersymmetric transformations: 
\begin{eqnarray}
\delta \bar{\psi}^{a}\,=\,\epsilon _{b}\,i\,D^{ab}\bar{A}\;\;\;\;
&,&\;\;\;\;\delta \psi _{b}\,=\,-\epsilon ^{c}\,i\,D_{bc}A\,,  \nonumber \\
\delta A\,=\,-\epsilon ^{a}\,\psi _{a}\;\;\;\; &,&\;\;\;\;\delta \bar{A}%
\,=\,-\epsilon ^{a}\,\bar{\psi}_{a}\,,
\end{eqnarray}

The Noether current associated to $N$=$1$--supersymmetry is now: 
\begin{eqnarray}
{J^{\mu }}_{c} &=&\frac{i}{2}\,\varepsilon ^{\mu \,\nu \,\rho }\,[\bar{\psi}%
^{a}\,(\gamma _{\rho })_{ac}\,\partial _{\nu }\,A\,+\psi ^{a}\,(\gamma
_{\rho })_{ac}\,\partial _{\nu }\,\bar{A}]\,-\,\frac{1}{2}\,(\psi
_{c}\,\partial ^{\mu }\,\bar{A}\,+\,\bar{\psi}_{c}\,\partial ^{\mu }\,A)\,+ 
\nonumber \\
&-&\frac{1}{2}\,(\bar{A}\,\partial ^{\mu }\,\psi _{c}\,+\,A\,\partial ^{\mu
}\,\bar{\psi}_{c})\,+\,\frac{i}{2}\,\varepsilon ^{\mu \,\nu \,\rho }\,(\bar{A%
}\,\partial _{\nu }\,\psi ^{a}\,+\,A\,\partial _{\nu }\,\bar{\psi}%
^{a})(\gamma _{\rho })_{ac}\,,
\end{eqnarray}
yielding the supercharge 
\begin{eqnarray}
Q_{c} &=&\int d^{2}\vec{x}{J^{0}}_{c}=  \nonumber \\
&=&\int d^{2}\vec{x}\frac{1}{2}\left\{ -\,(\psi _{c}\,\partial ^{0}\,\bar{A}%
\,+\,\bar{\psi}_{c}\,\partial ^{0}\,A)\,+\,i\,\varepsilon ^{0\,i\,j}\,[\bar{%
\psi}^{a}\,(\gamma _{j})_{ac}\,\partial _{i}\,A\,+\psi ^{a}\,(\gamma
_{j})_{ac}\,\partial _{i}\,\bar{A}]+\right.  \nonumber \\
&&-\,\left. (\bar{A}\,\partial ^{0}\,\psi _{c}\,+\,A\,\partial ^{0}\,\bar{%
\psi}_{c})\,+\,i\,\varepsilon ^{0\,i\,j}\,(\bar{A}\,\partial _{i}\,\psi
^{a}\,+\,A\,\partial _{i}\,\bar{\psi}^{a})(\gamma _{j})_{ac}\right\}
\label{cargascalgauge}
\end{eqnarray}

The canonical conjugate momenta that will be necessary for the supercharge
algebra are 
\begin{eqnarray}
\Pi_{\psi_d} \, = - i \bar{\psi}^a {(\gamma^0)_a}^d & , & \Pi_{\bar{\psi}_d}
\, = \, - \, i \psi^a {(\gamma^0)_a}^d \; ,  \nonumber \\
\Pi_A \, = \, (D^0 \bar{A}) & , & \Pi_{\bar{A}} \, = \, (D^0 A) \, ,
\end{eqnarray}
giving the canonical commutation and anticommutation relations 
\begin{eqnarray}
\{ \psi^d \, , \, \bar{\psi}^a \} \, = \, i \, (\gamma^0)^{ad} \, \delta^2 (%
\vec{x} - \vec{y}) \; \; \; & , & \; \; \; \{ \bar{\psi}^d \, , \, \psi^a \}
\, = \, i \, (\gamma^0)^{ad} \, \delta^2 (\vec{x} - \vec{y}) \, ,  \nonumber
\\
{[} A \, , \, D^0 \bar{A} {]} \, = \, \delta^2 (\vec{x} - \vec{y}) \; \; \;
& , & \; \; \; [ \bar{A} \, , \, D^0 A ] \, = \, \delta^2 (\vec{x} - \vec{y}%
) \, .
\end{eqnarray}

After a lengthy computation, using the $\gamma $--matrices Clifford algebra,
we reach the result 
\begin{equation}
\{Q_{a}\,,\,Q_{b}\}=-2\,i\,P^{\mu }(\gamma _{\mu })_{ab}\,,
\label{algscalgau}
\end{equation}
where the momentum operator $P^{\mu }$ appearing in the RHS includes now
contributions from the gauge field minimally coupled to matter through (\ref
{acaoscalgau}). Nevertheless, no term in the form of a central charge arises
from the action (\ref{acaoscalgau}); this means that the central charge
operator of eq. (\ref{mattercharge}) is not modified by the introduction of
the $U(1)$ gauge superfield. The role of the latter is to stabilize the
topological configurations associated to the action (\ref{acaosuper}) in the
form of vortex-like solitons, as already known from the works quoted in ref. 
\cite{rider}.

\section{The Supersymmetric Chern--Simons Term}

Now, we shall add a supersymmetric Chern--Simons (CS) term to the action eq.
(\ref{acaoscalgau})and we will verify how it modifies the supercharge
algebra. For this purpose, we begin with the (gauge--invariant) CS term in
superspace 
\begin{equation}
S_{CS}=\frac{M}{g^{2}}\,\int \,d^{3}x\,d^{2}\theta \;\Gamma ^{a}\,W_{a}\,,
\label{acaocs}
\end{equation}
where $M$ is a mass parameter and $g$ is the gauge coupling constant. In
components, using the {\it Wess--Zumino gauge}, the action eq. (\ref{acaocs}%
) leads to the expression 
\begin{equation}
S_{CS}\,=\,\frac{M}{g^{2}}\int d^{3}x\left[ i\,V^{ab}\,(\partial _{ac}\,{%
V^{c}}_{b})\,+\,4\,\lambda ^{2}\right] \,,  \label{acaocscomp}
\end{equation}
where the first term in the r.h.s. is the well--known CS term. Now,
including the term (\ref{acaocs}) in the action (\ref{acaoscalgau}), and
then taking into account the equations of motion for the $F$ , $\bar{F}$
(which are not affected by the CS term) and of the $\lambda _{a}$--field,
the complete Lagrangian reads: 
\begin{eqnarray}
{\it L}_{din} &=&\frac{i}{2}\bar{\psi}^{a}{(\gamma ^{\mu })_{a}}^{b}\partial
_{\mu }\psi _{b}+\frac{i}{2}\psi ^{a}{(\gamma ^{\mu })_{a}}^{b}\partial
_{\mu }\bar{\psi}_{b}-\frac{1}{2}(\partial ^{\mu }\bar{A})(\partial _{\mu
}A)+\frac{i}{2}(\partial ^{\mu }\bar{A})V_{\mu }A+  \nonumber \\
&&-\frac{i}{2}V^{\mu }\bar{A}(\partial _{\mu }A)+\frac{iM}{g^{2}}\epsilon
^{\mu \nu \rho }V_{\mu }\partial _{\nu }V_{\rho }\,,
\end{eqnarray}
with the ``on shell'' supersymmetric transformations: 
\begin{eqnarray}
\delta \bar{\psi}^{a}\,=\,i\epsilon _{b}D^{ab}\bar{A}\;\;\;\;
&,&\;\;\;\;\delta \psi _{b}\,=\,-i\epsilon ^{c}D_{bc}A\,,  \nonumber \\
\delta A\,=\,-\epsilon ^{a}\psi _{a}\;\;\;\; &,&\;\;\;\;\delta \bar{A}%
\,=\,-\epsilon ^{a}\bar{\psi}_{a}\,,  \nonumber \\
\delta V_{ab}\, &=&\,i\epsilon _{(a}\lambda _{b)}\,.  \label{susytransf}
\end{eqnarray}

Following again the procedure to read off the Noether current associated to
the transformations (\ref{susytransf}), it can be found out that the
contribution of the CS term yields: 
\begin{equation}
(J_{SCS})_{c}^{\mu }\,=\,\frac{i}{2}\,V^{\mu }\left( \psi _{c}\bar{A}-\bar{%
\psi}_{c}A\right) \,,
\end{equation}
with the supercharge 
\begin{equation}
(Q_{SCS})_{c}\,=\,\int d^{2}\vec{x}(J_{SCS})_{c}^{0}\,=\,\int d^{2}\vec{x}%
\left\{ \frac{i}{2}V^{0}\left( \psi _{c}\bar{A}-\bar{\psi}_{c}A\right)
\right\} \,,
\end{equation}
whence 
\begin{equation}
\{Q_{a}\,,Q_{b}\}_{SCS}\,=\,2\,i\,P^{\mu }[V^{0}](\gamma _{\mu })_{ab}\,.
\end{equation}
where $P^{\mu }[V^{0}]$ means a functional that depends only on the
time--component of the potential vector.

\smallskip

What we observe is that the $V^{0}$ potential field is completely eliminated
from the algebra, implying that the ``corrected'' Chern-Simons $T^{0\,\mu }$
component of energy-momentum tensor, defined as ``new'' $P^{\mu }$ becomes
independent on the potential gauge field. It is possible to say that the the
conjugate momenta of the $A$ and $\bar{A}$ fields are in fact ``corrected''
by the CS term to become $\Pi _{A}\,\alpha \,\partial ^{0}\bar{A}$ and $\Pi
_{\bar{A}}\,\alpha \,\partial ^{0}A$ . This indicates that the CS term plays
a role similar to a partial gauge fixing, eliminating one degree of freedom
of the gauge field, referring to the algebra.

\section{Conclusions}

The basic motivation of this paper was to analyze the 3-dimensional
counterpart of a well-known result by Olive and Witten \cite{witten},
namely, the appearance of a central charge in the algebra of simple
supersymmetry as originated from non-trivial topological field
configurations. To this aim we have analyzed the full supersymmetric model
in 3D. We have computed the Noether supersymmetric charges, and the
``improved'' energy-momentum tensor using the gravitation approach. From the
canonical commutations (and anti-commutations) relations of the superfields
we obtained the supercharge algebra. Here, with and without gauge fields, we
could conclude that vortex-like field configurations are responsible for a
central charge in the supersymmetry algebra, even in the case of a $N=1$%
-supersymmetry. It is worthwhile to mention the results obtained by Lee, Lee
and Weinberg \cite{lee}, where a central charge comes out in context of an $%
N=2$ extended supersymmetric model. We would like to point out that the
calculations of Section 3 recall that the Chern-Simons term for the gauge
field does not give contribution to the central charge appearing in the
algebra. In fact it could represent a sample of gauge fixing eliminating the
time direction of the vector potencial in the algebra. So the central charge
of this model arises exclusively from the matter sector and its existence to
the vortex configurations of the scalar fields. Clearly, the role of the
gauge fields is to render finite the vortex energy \cite{rider}. The
analysis of the BPS bounds and its consequences will appear in a forthcoming
paper.

Next, it might be of relevance to analyze the presence of central charges in
the models proposed by Dorey and Mavromatos \cite{dorey} to study $P,T$
conserving superconducting gauge models whenever the latter are
supersymmetrised. One could perhaps understand whether or not central
charges may be related to some physical aspects of superconductivity.

\vspace{.5cm}

{\bf Acknowledgements}: The author is grateful to Dr. J. A. Helay\"{e}l-Neto
for all the discussions and the critical comments on the original
manuscript; thanks are also due to Dr. O. Piguet for very pertinent and
helpful remarks, and to the colleagues at DCP-CBPF for the general
discussions. CAPES-Brasil is also acknowledged for the invaluable financial
help.


\begin{thebibliography}{99}
\bibitem{alguem}  {A. Linde, {\em {Rep. Progr. Phys.}} {\bf {42}} (1979) 389,%
} \newline
{S. Deser, R. Jackiw and S. Templeton, {\em {Phys. Rev. Lett.}} {\bf {48}}
(1982) 975 , \newline
S. Deser, {\em {Three Topics in Three Dimensions}}, in the proceedings of
the Trieste Conference on Supermenbranes and Physics in 2+1 Dimensions, eds.
M.J. Duff {\em {et al.}}, World Scientific (Singapore, July 1989)};

M.A.{De Andrade, O.M.Del Cima, L.P.Colatto, {\em Phys. Lett.} {\bf 370B}
(1996) 59},

L.P.{Colatto, M.A.De Andrade, O.M.Del Cima, D.H.T.Franco,J.A.
Helay\"{e}l-Neto, O. Piguet, }{\em Jour. Phys. }{\bf G24 }(1998) 1301,\ \ \ 

\bibitem{piguet}  F. Delduc, C. Lucchesi, O. Piguet and S.P. Sorella, {\em {%
Nucl.Phys.}} {\bf {B346}} (1990) 313 , \newline
A. Blasi, O. Piguet and S.P. Sorella, {\em {Nucl.Phys.}} {\bf {B356}} (1991)
154 , \newline
C. Lucchesi and O. Piguet, {\em {Nucl.Phys.}} {\bf {B381}} (1992) 281,

\ \ \ \ \ \ O.M.Del Cima....

\bibitem{alguemdois}  {J. Schonfeld, {\em {Nucl. Phys.}} {\bf {B185}} (1981)
157},

\bibitem{grisaru}  S. J. Gates, M. T. Grisaru, M. Ro\u {c}ek and W. Siegel, 
{\em Superspace}, \newline
Benjamin/Cummings Publishing Company (1983),

\bibitem{haag}  S. Coleman and J. Mandula, {\em Phys. Rev.} {\bf 159} (1967)
1251, \newline
R. Haag, J. T. Lopuszanski and M. Sohnius, {\em Nuc. Phys.} {\bf B88} (1975)
257;

\bibitem{jackiw}  R. Jackiw, {\em Field Theoretic Investigation in Current
Algebra}, {\em Current Algebra and Anomalies}, S. B. Treiman, R. Jackiw, B
Zumino and E. Witten (eds.), World Scientific Publishing Co. (1985),

\bibitem{witten}  E. Witten and D. Olive, {\em Phys. Lett.} {\bf B78} (1978)
97, \newline
P. Di Vecchia and S. Ferrara, {\em Nuc. Phys.} {\bf B130} (1977) 93, \newline
A. D'Adda and P. Di Vecchia, {\em Phys. Lett.} {\bf B73} (1978) 162,

\bibitem{rider}  Lewis H. Ryder, {\em Quantum Field Theory} - Chap. 9,
Cambridge University Press, 1985,

\bibitem{lee}  C. Lee, K. Lee and E. Weinberg, {\em Phys. Lett.} {\bf B243}
(1990) 105,

\bibitem{dorey}  N. Dorey and N. E. Mavromatos, {\em Nuc. Phys.} {\bf B386}
(1992) 614; \newline
{N. E. Mavromatos, {\em {Superconducting Gauge Theories in (2+1)-Dimensions}}%
, CERN preprint, CERN-TH.6331/91 (1991)}.
\end{thebibliography}
\end{document}